\documentclass{aastex62}

\usepackage{comment}
\begin{document}

\title{The Enhancement of Proton Stochastic Heating in the near-Sun Solar Wind}

\correspondingauthor{M. M. Martinovi\'c}
\email{mmartinovic@email.arizona.edu}

\author[0000-0002-7365-0472]{Mihailo M. Martinovi\'c}
\affiliation{Lunar and Planetary Laboratory, University of Arizona, Tucson, AZ 85721, USA.}
\affiliation{LESIA, Observatoire de Paris, Meudon, France.}

\author[0000-0001-6038-1923]{Kristopher G. Klein}
\affiliation{Lunar and Planetary Laboratory, University of Arizona, Tucson, AZ 85721, USA.}

\author[0000-0002-7077-930X]{Justin C. Kasper}
\affiliation{Climate and Space Sciences and Engineering, University of Michigan, Ann Arbor, MI 48109, USA}
\affiliation{Smithsonian Astrophysical Observatory, Cambridge, MA 02138 USA.}

\author[0000-0002-3520-4041]{Anthony W. Case}
\affiliation{Smithsonian Astrophysical Observatory, Cambridge, MA 02138 USA.}

\author[0000-0001-6095-2490]{Kelly E. Korreck}
\affiliation{Smithsonian Astrophysical Observatory, Cambridge, MA 02138 USA.}

\author{Davin Larson}
\affil{Space Sciences Laboratory, University of California, Berkeley, CA 94720-7450, USA}

\author[0000-0002-0396-0547]{Roberto Livi}
\affil{Space Sciences Laboratory, University of California, Berkeley, CA 94720-7450, USA}

\author[0000-0002-7728-0085]{Michael Stevens}
\affil{Smithsonian Astrophysical Observatory, Cambridge, MA 02138 USA.}

\author[0000-0002-7287-5098]{Phyllis Whittlesey}
\affil{Space Sciences Laboratory, University of California, Berkeley, CA 94720-7450, USA}

\author[0000-0003-4177-3328]{Benjamin D. G. Chandran}
\affil{Department of Physics \& Astronomy, University of New Hampshire, Durham, NH 03824, USA}
\affil{Space Science Center, University of New Hampshire, Durham, NH 03824, USA}

\author[0000-0001-6673-3432]{Ben ~L.~Alterman}
\affiliation{Climate and Space Sciences and Engineering, University of Michigan, Ann Arbor, MI 48109, USA}
\affiliation{Department of Applied Physics, University of Michigan, 450 Church St., Ann Arbor, MI 48109, USA}

\author[0000-0002-9954-4707]{Jia Huang}
\affiliation{Climate and Space Sciences and Engineering, University of Michigan, Ann Arbor, MI 48109, USA}

\author[0000-0003-4529-3620]{Christopher H. K. Chen}
\affil{School of Physics and Astronomy, Queen Mary University of London, London E1 4NS, UK}

\author[0000-0002-1989-3596]{Stuart D. Bale}
\affil{Space Sciences Laboratory, University of California, Berkeley, CA 94720-7450, USA}
\affil{School of Physics and Astronomy, Queen Mary University of London, London E1 4NS, UK}
\affil{Physics Department, University of California, Berkeley, CA 94720-7300, USA}
\affil{The Blackett Laboratory, Imperial College London, London, SW7 2AZ, UK}

\author[0000-0002-1573-7457]{Marc Pulupa}
\affil{Space Sciences Laboratory, University of California, Berkeley, CA 94720-7450, USA}

\author[0000-0003-1191-1558]{David M. Malaspina}
\affiliation{Laboratory for Atmospheric and Space Physics, University of Colorado, Boulder, CO, USA}

\author[0000-0002-0675-7907]{John W. Bonnell}
\affil{Space Sciences Laboratory, University of California, Berkeley, CA 94720-7450, USA}

\author[0000-0002-6938-0166]{Peter R. Harvey}
\affil{Space Sciences Laboratory, University of California, Berkeley, CA 94720-7450, USA}

\author[0000-0003-0420-3633]{Keith Goetz}
\affil{School of Physics and Astronomy, University of Minnesota, Minneapolis, MN 55455, USA}

\author[0000-0002-4401-0943]{Thierry {Dudok de Wit}}
\affil{LPC2E, CNRS and University of Orl\'eans, Orl\'eans, France}

\author[0000-0003-3112-4201]{Robert J. MacDowall}
\affil{Solar System Exploration Division, NASA/Goddard Space Flight Center, Greenbelt, MD, 20771, USA}

\begin{abstract}

Stochastic heating is a non-linear heating mechanism driven by the violation of magnetic moment invariance due to large-amplitude turbulent fluctuations producing diffusion of ions towards higher kinetic energies in the direction perpendicular to the magnetic field. It is frequently invoked as a mechanism responsible for the heating of ions in the solar wind. Here, we quantify for the first time the proton stochastic heating rate $Q_\perp$ at radial distances from the Sun as close as $0.16$ au, using measurements from the first two Parker Solar Probe encounters. Our results for both the amplitude and radial trend of the heating rate, $Q_\perp \propto r^{-2.5}$, agree with previous results based on the Helios data set at heliocentric distances from 0.3 to 0.9 au. Also in agreement with previous results, $Q_\perp$ is significantly larger in the fast solar wind than in the slow solar wind. We identify the tendency in fast solar wind for cuts of the core proton velocity distribution transverse to the magnetic field to exhibit a flat-top shape. The observed distribution agrees with previous theoretical predictions for fast solar wind where stochastic heating is the dominant heating mechanism. 
\end{abstract}

\keywords{space plasmas, interplanetary turbulence, solar wind}

%% Start the main body of the article. If no sections in the 
%% research note leave the \section call blank to make the title.
\section{Introduction} 
\label{sec:Intro}

The first measurements of the solar wind throughout the inner heliosphere made by the two Helios spacecraft found that the ion temperature decreases with radial distance $r$ from the Sun is slower than expected from adiabatic expansion \citep{Marsch_1982,Kohl_1998}. Identifying what mechanisms drive this apparent heating of the solar wind thus became a central problem of heliophysics. As the standard fluid \citep{Parker_1958,Sturrock_1966,Hartle_1970,Wolff_1971} or exospheric \citep{Jockers_1970,Lemaire_1971} models were not able to explain the measured temperature profiles without adding ad-hoc sources of energy, the community concentrated its efforts on investigating different heating mechanisms that might produce the observed temperature profile.

After Alfv{\' e}n waves (AWs) were observed in situ in the solar wind \citep{Coleman_1968,Belcher_1969,Belcher_1971}\footnote{AWs were also observed by remote infrared measurements in the solar corona \citep{Tomczyk_2007_Sci}}, a great deal of attention was devoted to modeling Alfv{\' e}nic turbulence. It was identified that the majority of proton heating by Alfv{\' e}nic fluctuations happens at the scale of the proton gyroradius $\rho_p$ \citep{Quataert_1998} and/or the proton inertial length $d_p$ \citep{Leamon_2000,Galtier_2006}, both being of the same order of magnitude in the solar wind. 
One of the proposed mechanisms that leads to strong perpendicular heating of both protons and heavy ions, as is observed in both remote \citep{Kohl_1998,Cranmer_1999} and in situ
\citep{Marsch_1982,Stansby_2018_SolPh,Stansby_2019_MNRAS,Perrone_2019_MNRAS} measurements, is ion cyclotron wave damping \citep{Isenberg_1983_JGR,Hollweg_1999JGR_1,Hollweg_1999JGR_2,Kasper_2013}. However, this process is quenched for the so-called critically balanced turbulence \citep{Goldreich_1995,Cho_2003_MNRAS,Schekochihin_2009}. In critical balance, the 'information' transfer through the system carried by Alfv{\' e}n waves enforces $k_\perp \delta v \sim k_{\parallel} v_A$ \citep{Howes_2015_RSPTA,Mallet_2015}, where $\delta v$ are velocity fluctuations at scale $\rho_p$ and $v_A = B / \sqrt{\mu_0 n_p m_p}$ is the Alfv{\' e}n velocity ($\mu_0$ is the permeability of vacuum, $n_p$ is the total proton density and $m_p$ is the proton mass), while $k_{\parallel}$ and $k_\perp$ are the parallel and perpendicular components of the wavevector $\mathbf{k}$ with respect to the magnetic field $\mathbf{B}$. Such a system has a turbulent power spectrum that is highly anisotropic, with more power at wavevectors with $k_\perp \gg k_{\parallel}$. An increasing set of solar wind observations \citep{Bale_2005,Chen_2010,Salem_2012_ApJ,Chen_2013_PhRvL,Safrankova_2019_ApJ} suggests that, as turbulent fluctuations approach ion scale $k_\perp \rho_p \sim 1$, the anisotropic turbulent AW cascade transforms into a kinetic Alfv{\' e}n wave (KAW) cascade due to finite Larmor radius effects \citep{Howes_2006_ApJ,Howes_2008b}. At these wavevectors, KAWs have frequencies lower than the ion cyclotron frequency and are not able to significantly damp via the cyclotron resonance with thermal ions.

Other proposed linear mechanisms, Landau and transit time damping, which both couple to the Landau resonance, depend on the ratio of thermal to magnetic pressures $\beta = v_t^2 / v_A^2$, where $v_t = \sqrt{2 k_b T_p / m_p}$ is the proton thermal velocity, with $k_b$ and $T_p$ being the Boltzmann constant and proton temperature, respectively. For $\beta < 1$, the Alfv\'en wave phase speed $\omega/k_\parallel \sim v_A$ is much greater than the bulk of the proton velocities, quenching the Landau resonance at ion scales. For $\beta \geq 1$, the Alfv\'en wave phase speed is comparable to typical proton velocities, which combined with the finite parallel electric field $\mathbf{E_{||}}$ produced by gyroscale KAWs enables Landau damping to heat protons. This mechanism has been observed in gyrokinetic simulations \citep{TenBarge_2013,Klein_2017_JPlPh,Howes_2018} and measured in the magnetosheath \citep{Chen_2019}, but has not yet been directly measured in the undisturbed solar wind, potentially due to current instrumental limitations. The difficulty in observing Landau damping and the concurrent radial heating profile is that it preferentially produces higher parallel temperatures, in conflict with the ion temperature anisotropy $T_\perp/T_\parallel >1$ observed in the corona and the inner heliosphere for $r < 0.5$ au.
We point the reader to various review articles containing descriptions of solar wind turbulence properties \citep{Chen_2016_JPlPh}, heating mechanisms proposed by the community \citep{Cranmer_2015}, as well as the historical development of the field through theory and observations \citep{Schekochihin_2019_JPP}. 

In this article, we focus on one nonlinear heating mechanism: the stochastic heating (SH) of protons. When a proton gyrates about some magnetic field, its magnetic moment $\mu_m = m_p v_\perp^2/2 B$ ($v_\perp$ being the proton velocity perpendicular to $\mathbf{B}$) is conserved, as long as changes in the magnetic field are sufficiently slow. However, if turbulent fluctuations at the proton gyroradius scale are sufficiently large, they will lead to the violation of the magnetic moment conservation by causing ion orbits to become non-periodic in the plane perpendicular to \textbf{B}. The change in an individual particle's energy can be either positive or negative but, as  long  as  the  particle distribution is monotonically decreasing towards higher kinetic energies, the net transfer of energy is from the electromagnetic fields to the particle distribution, leading to ion heating \citep{Chandran_2010}. This phenomenon was first observed in tokamaks \citep{McChesney_1987_PhRvL,McChesney_1991_PhFlB}, and then followed up by comprehensive theoretical and observational work for the case of SH due to electrostatic waves \citep{Ping-Kun_1996_PhPl,Chen_2001_PhPl}, as well as for specific cases of cometary environments \citep{Karimabadi_1994}, nonlinear structures \citep{Stasiewicz_2000_PhST}, the magnetopause \citep{Johnson_2001_GeoRL}, aurora \citep{Chaston_2004_JGRA}, laboratory reversed-field pinch \citep{Fiksel_2009_PhRvL} and, finally, free solar wind \citep{Voitenko_2004_ApJ,Chandran_2010,Chandran_2010a}.

A theory that quantifies the perpendicular SH rate $Q_\perp$ in the solar wind was developed by \citet{Chandran_2010} for low-$\beta$ plasma streams. They show that $Q_\perp$ crucially depends on the ratio between the amplitude of the turbulent velocity fluctuations near the proton gyroscale and perpendicular thermal velocity

\begin{equation}
\epsilon = \frac{\delta v}{v_{t\perp}}
\label{eq:Epsilon}
\end{equation}

\noindent and is given by

\begin{equation}
Q_\perp = \frac{c_1 (\delta v)^3}{\rho_p} \exp \left[-\frac{c_2}{\epsilon}\right].
\label{eq:Heating_Rate_low_beta}
\end{equation}
where $\rho_p = m_p v_{t\perp} / e_c B$ and $e_c$ is the elementary charge.

Equation \ref{eq:Heating_Rate_low_beta} was further tested using Helios observations at three radial distances \citep{Bourouaine_2013} and in RMHD simulations \citep{Xia_2013}. These tests indicate that SH might be very important for the solar wind heating, especially when other mechanisms are predicted to be suppressed. However, $Q_\perp$ strongly depends on the values of order unity constants $c_1$ and $c_2$ used in the model, bringing considerable uncertainties into the calculated heating rates. In a former study, \citet{Martinovic_2019_ApJ} (hereafter MKB19) processed the entire Helios 1 and 2 mission data sets and obtained reliable radial trends of $Q_\perp \sim r^{-3.1}$ for the fast ($v_{sw} > 600$ km/s) and $Q_\perp \sim r^{-0.6}$ for the slow ($v_{sw} < 400$ km/s) solar wind as well as order of magnitude estimates for the average value of $Q_\perp$. A variety of possible values for $c_1$ and $c_2$ was considered, as well the total fraction of ion heating associated with SH. 
Due to limits of the Helios instruments and the radial extent of the mission, 
an improved survey to determine the 
relative importance of SH is needed, and the Parker Solar Probe (PSP) \citep{Fox_2016} provides such measurements.

In addition to calculating $Q_\perp$, one can determine the effect of stochastic heating on the proton VDF shape. Towards this end, \citet{Klein_2016_ApJ} solved the gyroaveraged kinetic equation \citep{Kulsrud_1983_bpp} of a reduced proton VDF perpendicular to $\mathbf{B}$ for a low-$\beta$ plasma, using the diffusion coefficient given by \citet{Chandran_2010} to describe the influence of strong SH. They found that the reduced VDF evolves towards a flat-top shaped modified Moyal distribution, with a decreased value of excess kurtosis $\kappa$ of approximately $-0.8$ compared to the Maxwellian value of $0$. This structure can be used as evidence for the action of SH. Recent simulation work \citep{Isenberg_2019_ApJ} suggests that extreme flat-top distribution shapes might be suppressed in plasma with significant temperature anisotropies $T_\perp \gg T_{||}$ due to rise of the ion-cyclotron instability.
Studies of ion VDFs in Alfv\'enic turbulence \citep{Arzamasskiy_2019} produce a similar perpendicular flat-top distribution, although other structures arise at high energies and in parallel direction, likely due to ion cyclotron heating of suprathermal particles and pitch angle scattering at $v \sim v_t$.
As the PSP Solar Probe Cup (SPC) directly measures reduced VDFs, we were able to obtain the distribution shape for about 3.5 million measurements. Details of measuring the 
VDFs and other data processing are given in Section \ref{sec:Method}, while results are summarized in Section \ref{sec:Results}.
We examined the width of the 'core' part of the reduced VDFs, which corresponds to thermal protons. In the fast solar wind, the reduced VDFs are, on average, significantly wider compared to a Maxwellian with equivalent temperature when acquired in the perpendicular direction. A similar trend is not present in the slow wind, which suggests that SH is much stronger in the fast wind. This result is consistent with the findings of MKB19.

Discussion of the results is found in Section \ref{sec:Discussion} including a comparison of the $Q_\perp$ calculations with a simple model for the total solar wind heating rate in the perpendicular direction. We find that $Q_\perp$ measured at $r \leq 0.2$ au is comparable to the total heating rate of protons, while it becomes less significant at larger radial distances. This result is in agreement with strong SH radial trends found in MKB19 and is partially confirmed by measured flat-top VDFs in the fast solar wind. 

\section{Method} 
\label{sec:Method}

We characterize the properties of SH and its importance in solar wind heating using three complementary methods. First, we calculate the SH rate based on theoretical expressions provided by \citet{Chandran_2010} for low-$\beta$ plasma and \citet{Hoppock_2019} for all values of $\beta$. Second, following predictions by \citet{Klein_2016_ApJ}, we examine the reduced VDFs, looking for a flat-top shape in the plasma streams where SH is expected to be dominant. Third, we compare calculated SH rates with a simple model used by \citet{Bourouaine_2013} to estimate the total perpendicular heating rate of the solar wind (the term 'total' refers to the difference between measurements and an assumed adiabatic expansion).

The plasma measurements are acquired from the PSP SWEAP (Solar Wind Electrons, Protons and Alphas) \citep{Kasper_2015} instrument suite. Proton VDF moments are used, including proton density $n_p$, bulk velocity $\mathbf{v_{sw}}$ and temperature $T_p$ as measured by the SPC in the 'peak tracking' operation regime \citep{Case_2019_SPC}. The measurements flagged by the instrument team as potentially unreliable are disregarded. Temperature anisotropy is measured by \citet{Huang:prep}. The method of determining parallel and perpendicular temperatures relies on observing VDFs with different angles with respect to fluctuating $\mathbf{B}$ over short time intervals, for which the temperature variation is considered to be small. Magnetic field vector $\mathbf{B}$ measurements are provided by the PSP Fields instrument team \citep{Bale_2016}.

\begin{figure}
\centering
\includegraphics[width=0.6\textwidth]{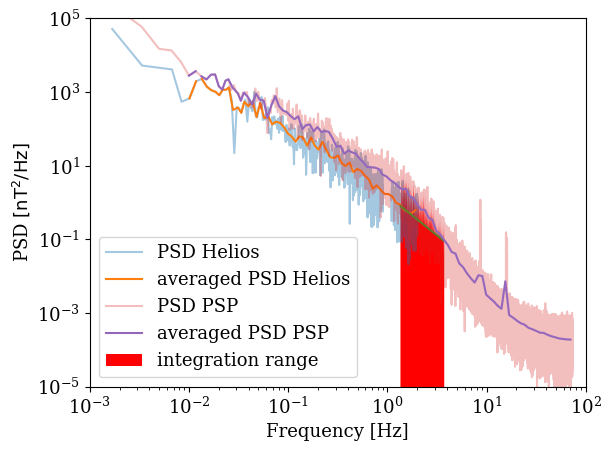}
\caption{Example of a processed 10-minute power spectrum (blue line) from Helios, observed between 03:50 and 04:00 on April 14, 1976 (reproduced from MKB19) and from PSP, observed between 03:30 and 03:40 on November 4, 2018 (red line). The frequency range is divided into logarithmically spaced regions and averaged within each of these regions (orange and violet solid lines). The spectral range observed in this work is shaded in red, matches within $2\%$ for the two intervals shown, and is below the instrumental spike at $f \approx 7.5$ Hz.}
\label{fig:PSD_Example}
\end{figure}

The method of SH rate calculation for low-$\beta$ plasma in this work is similar to the one applied in MKB19, and will be only briefly described here. As we are unable to directly measure the velocity fluctuations $\delta v$\footnote{A SPC operation mode that allows the study of high-cadence velocity fluctuations, called Flux Angle Mode, was active during a few limited time windows within the first two encounters. This mode, which samples a single energy channel very rapidly \citep{Vech:prep}, could potentially be used for verification of Equation \ref{eq:delta_v} in future work and the direct evaluation of $Q_\perp$.}, we first estimate the magnetic field fluctuations at the convective gyroscale
\begin{equation}
    \delta B = \Bigg[\frac{\pi}{C_0(n_{s})}  \int_{e^{-0.5} f_p}^{e^{0.5} f_p} P_B(f) df \Bigg]^{1/2} \label{eq:delta_B}
\end{equation}
where $P_B(f)$ is the level of a 10-minute magnetic field power spectrum, $f_p = v_{sw} \sin{\theta}/2\pi\rho_p$ is the convected gyrofrequency\footnote{The expresson for the convective gyrofrequency in MKB contains a typographical error; a factor of $2\pi$ is missing from the denominator. Also, Equation 8 of MKB19 should be identical to Equation \ref{eq:delta_B}, while that version has the factor $f_p$ inside the exponential factor in the integral boundaries. Neither of these typographical errors affect the contents presented in that article.} - the inverse of the time needed for the solar wind to advect a structure the size of a proton gyroradius and $\theta$ is the angle between $\mathbf{v_{sw}}$ and \textbf{B}. The magnetic field trace power spectrum $P_B(f)$ is calculated as a sum of fast Fourier transforms of each of the three components (red line on Figure \ref{fig:PSD_Example}). It is divided into 100 logarithmically spaced regions and averaged within each of these regions (violet solid line). Instrumental spikes visible on Figure \ref{fig:PSD_Example} originate from the spacecraft reaction wheels and have amplitudes below the measured signal for the spectral range of interest in this work. The four reaction wheels have magnetic signatures that correspond to their spin rate. Each wheel spins at an independent rate anywhere from a fraction of a Hz to several tens of Hz. The wheel rates generally drift in frequency slowly (over hours), but can vary abruptly in the case of a momentum dump \citep{Bowen:prep}. Finally, we perform a linear fit of the spectrum on a logarithmic scale in the range $[e^{-0.5} f_p, e^{0.5} f_p]$ to obtain the slope $n_s$ and geometrical factor \citep{Bourouaine_2013,Vech_2017}
\begin{equation}
C_0(n_{s}) = \frac{\pi^{0.5} \Gamma\big[{\frac{n_{s}}{2}}\big]}{2 \Gamma\big[{\frac{n_{s}+1}{2}}\big]}.
\label{eq:C_0}
\end{equation}
We assume that the fluctuations are Alfv{\' e}nic, and therefore relate
\begin{equation}
\delta v = \frac{\sigma v_A \delta B}{B}
\label{eq:delta_v}
\end{equation}
where $\sigma = 1.19$ is a dimensionless constant used by \citet{Chandran_2010} for a spectrum of randomly phased kinetic Alfv{\'e}n waves with $k_\perp \rho_\perp \sim 1$. Finally, the SH rate in the perperndicular direction is calculated using Equations \ref{eq:Epsilon} and \ref{eq:Heating_Rate_low_beta}, for cases of the plasmas with $\beta_{||} \leq 0.3$, which is equivalent to the analysis done on the Helios data set in MKB19.

%Here, we briefly overview the theoretical concept. The perpendicular heating can be represented as temporal variation of the system Hamiltonian, which has two components. The first term is determined by the electrostatic potential and is important in low-$\beta$ plasma, leading to Equation \ref{eq:Heating_Rate_low_beta}. Second term is determined by the vector potential and is important at moderate-high $\beta$ values. Contributions from both terms are combined by \citet{Hoppock_2019}}

Recent results by \citet{Hoppock_2019} give estimates for the SH rate for plasma streams where the above low-$\beta$ condition is not satisfied, with the general expression
\begin{equation}
Q_\perp = v_A^2 \Omega \Bigg( \sigma_1 \delta^3 \exp \left[-\frac{\sigma_2}{\delta}\right] + \frac{\sigma^3 c_1 \delta^3}{\beta^{1/2}} \exp \left[-\frac{c_2 \beta^{1/2}}{ \sigma \delta}\right] \Bigg)
\label{eq:Heating_Rate}
\end{equation}
where $\sigma_1$ and $\sigma_2$ are order unity constants and
\begin{equation}
\delta = \frac{\delta B}{B}.
\label{eq:delta}
\end{equation}
Exponential suppression factors in both Equations \ref{eq:Heating_Rate_low_beta} and \ref{eq:Heating_Rate} are introduced to account for property of the Hamiltonian variations (for $\epsilon << 1$) to be correlated and highly reversible over long time intervals to the leading order of $\epsilon$, and thus not contributing to perpendicular heating (see Appendix of \citet{Chandran_2010}). These factors are used for all of the measurements in this work. In previous work, we find $\epsilon < 0.1$ without any significant radial trend (MKB19). 

Here, it is important to emphasize two differences compared to the method used in MKB19. First, due to poor time resolution ($\Delta t = 0.25 - 0.5$ s) of the Helios E2 magnetometer \citep{Musmann_1975}, the power spectrum was available only down to 1 or 2 Hz (blue and orange line on Figure \ref{fig:PSD_Example}). Therefore, we performed a logarithmic fit of $P_B(f)$ and then extrapolated it as a power law if the convective gyroscale was out of measured range. On PSP Fields, magnetic field measurements with cadence as fast as 1/256 NYs are available, making the convected gyroscale range directly measurable, as shown by the red shaded area on Figure \ref{fig:PSD_Example}. NYs stands for New York second ($1 \mathrm{NYs} \approx 0.874$ s), a time unit used by Fields in order to synchronize with the PSP master clock \citep{Bale_2016}. For $r>0.25$ au we show no results as the Fields magnetometer resolution significantly decreases\footnote{Outside of the encounter 1, the Fields magnetometer resolution drops to $\Delta t \approx 0.43s$, while outside of encounter 2 we have $\Delta t \approx 0.1$. This value sets Nyquist frequency just below 5 Hz and could be, with some improvement of the method (i.e. more sophisticated treatment of the spectral break), used in our work. However, for significant time periods magnetometer data is missing,  SPC measurements have resolution comparable to the used time window (10 minutes) or are flagged by the instrument team as unreliable, so a potential upgrade of our processing methods in order to cover this very limited set of measurements is left for future projects.}. Second, finding an adequate PSD power law $n_s$ in the range of interest on Helios required estimation of the so-called break point, where the power spectrum transitions from the inertial range (with approximately a $f^{-5/3}$ slope) to the steeper dissipation range of frequencies. The high cadence and low noise floor of PSP measurements enable us to see the fine structure of the spectrum around the break. As its exact form is still widely debated in the literature \citep{Stawicki_2001_JGR,Markovskii_2008_ApJ,Perri_2010_ApJ,He_2012_ApJ,Bourouaine_2012_ApJ,Vech_2018b}, we refrain from establishing specific criteria for its localization, but rather perform a logarithmic fit over the entire integration range to obtain $n_s$. This approach introduces uncertainty of up to $5\%$ into the value of $C_0$ in Equation \ref{eq:C_0}, which is less than the uncertainties produced by other factors while calculating the SH rate.

SPC is a Faraday cup (FC) utilizing four $90^o$ wide plates oriented parallel to and not protected by PSP's heat shield \citep{Case_2019_SPC}. This specific design enables SPC to collect the bulk of the incoming solar wind flow\footnote{For more detailed information on functioning of a Faraday Cup, see \citep{Kasper_2006_JGRA} and Section 2.2.1 of \citet{Verscharen_2019}.}. The instrument data set provides the differential charge flux density in the direction perpendicular to the shield, that is, the number of elementary charges per unit time per unit of radially oriented area. The detector voltage ranges within $V = 0.3 - 4$ kV, with most of the measurements having $dV/V \sim 9\%$ resolution. This FC measures the reduced solar wind VDF in the direction perpendicular to the spacecraft heat shield with resolution of the order of $dv / v \sim dV / V$ and proton velocities ranging in $m_p v^2 = 2 e_c V$. The detailed procedure of calculating VDFs is described in Appendix A of \citet{Case_2019_SPC}.

The reduced VDFs are further processed in the following way. First, we detect the maximum of a VDF and its equivalent velocity channel $v_{sw_\mathrm{peak}}$ and shift this value to zero. Starting from the shifted zero, we find the velocity derivative of the shifted VDF $f_s(v)$ in both directions. The first measurement that satisfies
\begin{equation}
v \frac{d f_s(v)}{dv} > 0
\label{eq:VDF_Noise}
\end{equation}
is interpreted as a sign of noise and further data points are disregarded. We note that the criteria given in Equation \ref{eq:VDF_Noise} can also be triggered by dense proton beam \citep{Alterman:prep} or $\alpha$ particle populations, but we focus on thermal protons in this study and filter out these cases as well.

Finally, in order to search for a trend of flat-top shaped VDF, we observe the width of the VDF at $80\%$ of its maximum. For this purpose, we take into account only distributions that have measured values below the $80\%$ threshold both above and below $v_{sw_\mathrm{peak}}$ after the noise is removed. In order to find a better estimate of this value, we over-sample shifted VDFs down to 2 km/s velocity resolution.

During the first two encounters SPC has measured a total of 5,167,255 ion energy distributions. After excluding the data from the calibration, 'full scan' and 'flux angle' instrument modes \citep{Case_2019_SPC} and flagging potentially unreliable results, 4,669,965 ($90.4\%$) measurements are ratified by the instrument team. Out of this selection, after applying our criteria, we use the total of 3,424,674 ($73.3\%$) VDFs. Of these, 2,508,153 ($73.2\%$) are measured in the slow ($v_{sw} < 350$ km/s), 910,068 ($26.6\%$) in the intermediate (350 km/s $< v_{sw} < 550$ km/s) and 6,453 ($0.2\%$) in the fast ($v_{sw} > 550$ km/s) solar wind.

The empirical total solar wind heating rate in the perpendicular direction is calculated from the expression used by \citet{Bourouaine_2013}
\begin{equation}
Q_{\perp emp} = B v_{sw} \frac{d}{dr}\Bigg( \frac{k_b T_\perp}{m_p B} \Bigg)
\label{eq:Q_emp}
\end{equation}
This formula is obtained by multiplying the MHD equations \citep{Kulsrud_1983_bpp} with $v_\perp^2$ and integrating over velocity to find the heating rate in the steady state (see \citet{Chandran_2011} for more details). To find the derivative in Equation \ref{eq:Q_emp}, we assume that the quantity in parenthesis has a power law dependence on radial distance. In order to avoid well known effect of solar wind speed and temperature correlation that changes with $r$ \citep{Marsch_1982,Perrone_2019_MNRAS}, we separate our measurements at $r<0.25$ au into four 0.0225 au wide radial bins, with each of the bins divided into three ranges by $v_{sw}$: $(200-350)$ km/s, $(350-550)$ km/s and $(550-800)$ km/s, representing the slow, intermediate and fast solar wind. Inside each of the 12 bins, we perform a linear fit of $\mathrm{lg}(x)\big(\mathrm{lg}(r)\big)$, where $x=k_b T_\perp/m_p B$ and lg is the base 10 logarithm. The relation $x = 10^{a_0} r^{a_1}$, with $a_0$ and $a_1$ being the linear fit parameters for each bin, allows us to calculate
\begin{equation}
\frac{dx}{dr} = \frac{x(r_{max}) - x(r_{min})}{r_{max} - r_{min}}
\label{eq:Numeric_Derivative}
\end{equation}
where $r_{min}$ and $r_{max}$ are the minimal and maximal radial distance within a bin. We have also tested an alternative expression to Equation \ref{eq:Numeric_Derivative}, where we find the analytic derivative $dx/dr = 10^{a_0} a_1 \bar{r}^{a_1-1}$, with $\bar{r} = (r_{max} - r_{min}) / 2$. This approach provides results of the same order of magnitude. For some of the bins, we obtain $Q_{\perp emp} < 0$, which indicates perpendicular cooling. During these periods, we observe significant variations of the VDF moments, so it is likely these results are a consequence of sampling different solar wind streams during the interval rather than a general radial trend; therefore we disregard these values. More sophisticated evaluations of $Q_{\perp emp}$ will be used in future work to account for such changes. If future studies of the young solar wind, supported by better statistics, show perpendicular cooling at specific conditions, this will impose as an important topic for future work.

%This formula relies on previous two-fluid multiple moment models \citep{Svendsen_2001} and shearing box simulations of the turbulence driven by the magnetorotational instability \citep{Sharma_2006}. The same expression can be derived form more sophisticated heating models \citep{Hellinger_2011,Hellinger_2013} by neglecting the flux terms.

\section{Results} 
\label{sec:Results}

\begin{figure*}
\gridline{\fig{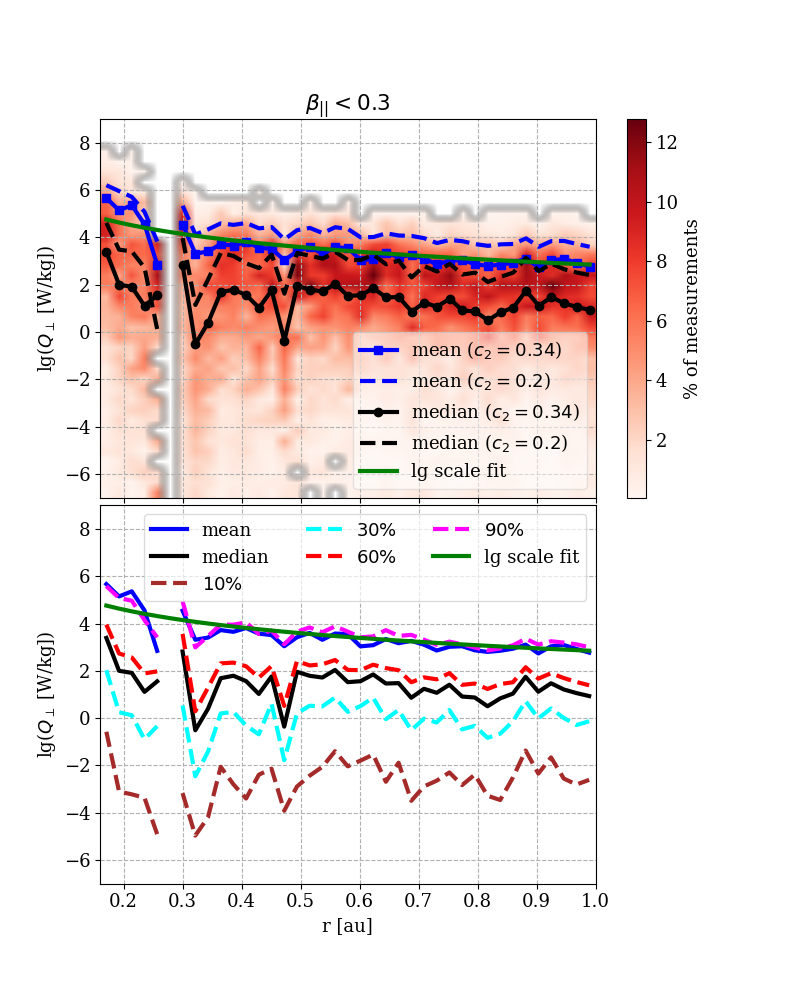}{0.49\textwidth}{(a)}
          \fig{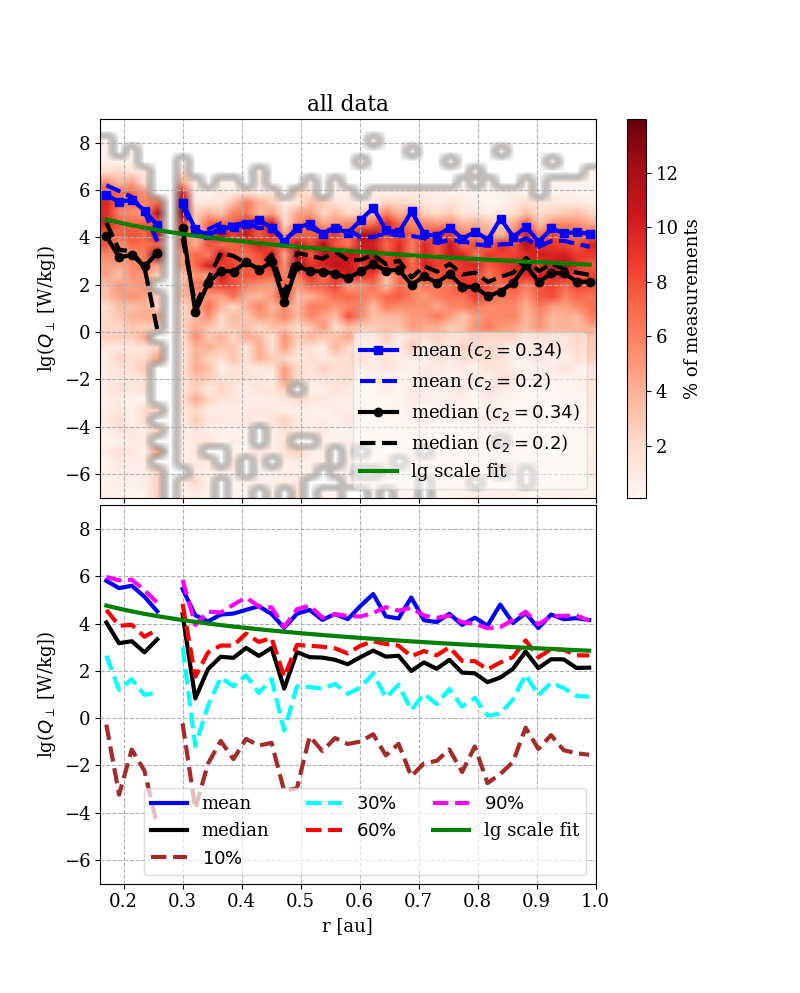}{0.49\textwidth}{(b)}
          }
    \caption{Perpendicular solar wind heating rate due to SH calculated in low-$\beta$ plasma streams using Equation \ref{eq:Heating_Rate_low_beta} (a) and using the general expression given in Equation \ref{eq:Heating_Rate} (b) throughout two PSP encounters (for $r \leq 0.25$ au) and entire Helios 1 and 2 missions (for $r \geq 0.29$ au) using $c_2 = 0.34$. Results that include high $\beta$ streams (Equation \ref{eq:Heating_Rate}) are significantly higher for both mean and median. The radial trend $Q_\perp \sim r^{-2.5}$, fitted from Helios observations} by MKB19, is shown by the green solid line, matching well with the results on panel (b), while on panel (a) we observe slightly stronger radial trend closer to the Sun. Results using $c_2 = 0.2$ (dashed lines) follow similar trends but with notably higher amplitudes. Data in upper panels shown is column normalized inside each radial bin, with results inside each column being uniformly binned in the logarithmic space. Lower panels show percentiles of the $Q_\perp$ distribution, which is highly skewed towards higher values, even in logarithmic space. Only $\sim 10-15\%$ of measured values are above average and contribute the most to the total SH in the solar wind.
    \label{fig:Q}
\end{figure*}

Merged values of the estimated SH rates for both Helios and PSP are shown on Figure \ref{fig:Q} for low-$\beta$ streams only (a) and for the entire PSP and Helios data set (b). As the measured $Q_\perp$ is very sensitive to the amplitude of turbulent fluctuations $\delta B$ (Equations \ref{eq:Heating_Rate_low_beta} and \ref{eq:Heating_Rate}), we bin the SH rate results evenly in the logarithmic space. The green line on both plots shows radial trend of $Q_\perp \sim r^{-2.5}$, obtained from Helios observations in MKB19. This trend is preserved, within the measurement uncertainties, in PSP results. The dip in the Helios results between 0.3 and 0.55 au is due to limitations of the Helios E2 magnetometer, discussed in detail in MKB19. The model constants used are obtained from test particle simulations \citep{Chandran_2010,Hoppock_2019}: $c_1 = 0.75$, $c_2=0.34$, $\sigma_1 = 5$, $\sigma_2 = 0.21$. The values of $c_1$ and $c_2$ are higher than the ones found by reduced RMHD simulations \citep{Xia_2013}, where $c_2$ can be as low as 0.15. As described in MKB19, decreasing $c_2$ by a factor of 2.5 increases $Q_\perp$ (in low-$\beta$ plasma) by one and a half orders of magnitude. Dashed lines show mean and median values with $c_2 = 0.2$, illustrating that the radial trends on panel (a) are almost completely preserved. The major feature of the results shown on panel (b) is that they are, on average, a factor 2-4 higher compared to low-$\beta$ values for assumed $c_2 = 0.34$, while for $c_2 = 0.2$ low-$\beta$ contribution becomes comparable to the high-$\beta$ component. However, it is important to note that the SH rates calculated for $\beta \geq 1$ using Equation \ref{eq:Heating_Rate} also assume purely Alfv\'enic fluctuations. As discussed in Section \ref{sec:Discussion}, this assumption might not be fully justified in high-$\beta$ solar wind, leading to potential overestimation of $Q_\perp$. 

\begin{figure*}
    \includegraphics[width=\textwidth]{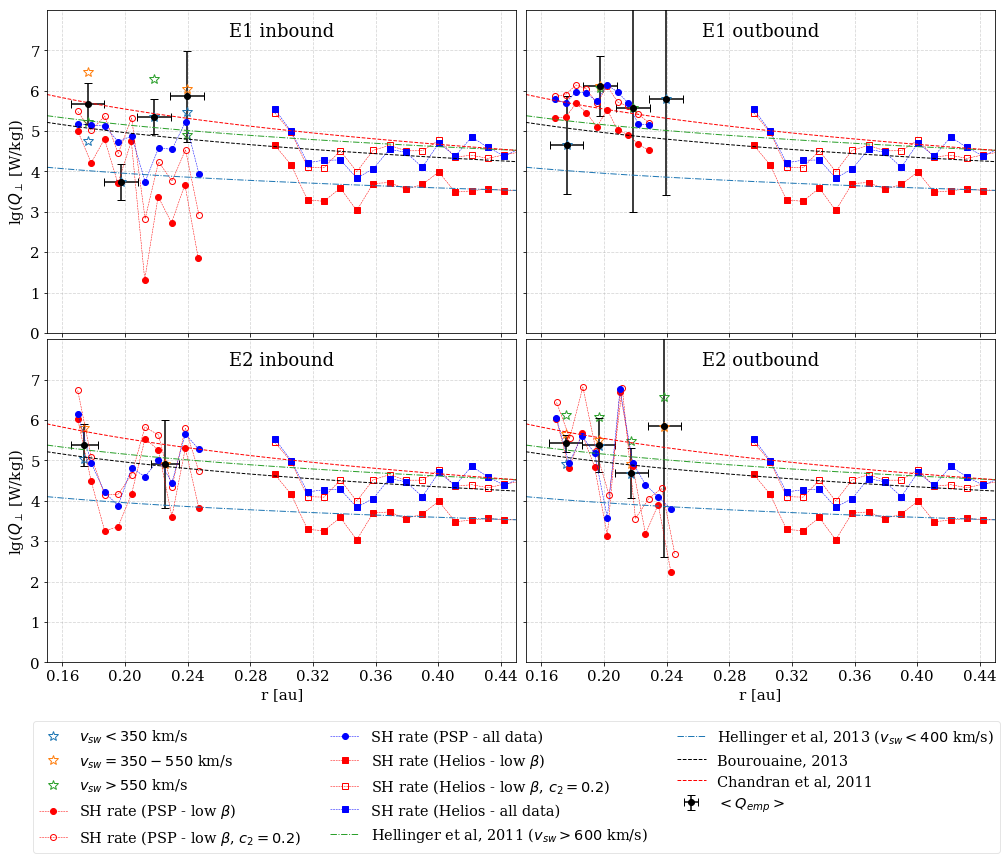}
    \caption{Estimated turbulent heating rates derived from MHD plasma parameters, following \citet{Bourouaine_2013} for PSP encounters 1 and 2, inbound and outbound. Black dots represent weighted averages of the total empirical perpendicular heating rates, while colored stars show the same result binned into slow, intermediate and fast solar wind. Estimated values of SH for low-$\beta$ plasma (filled red dots) and for all $\beta$ values (filled blue dots) are of the same order of magnitude as the total estimated heating rate for PSP measurements (except during E1 inbound phase), while it is not the case for Helios measurements (squares). Also, both measurement sets are comparable to theoretical models \citep{Chandran_2011,Bourouaine_2013}. SH rate values measured by PSP are consistently enhanced  for $c_2 = 0.2$ (hollow circles).}
    \label{fig:Empirical_Heating_Rates}
\end{figure*}

\begin{figure*}
\gridline{\fig{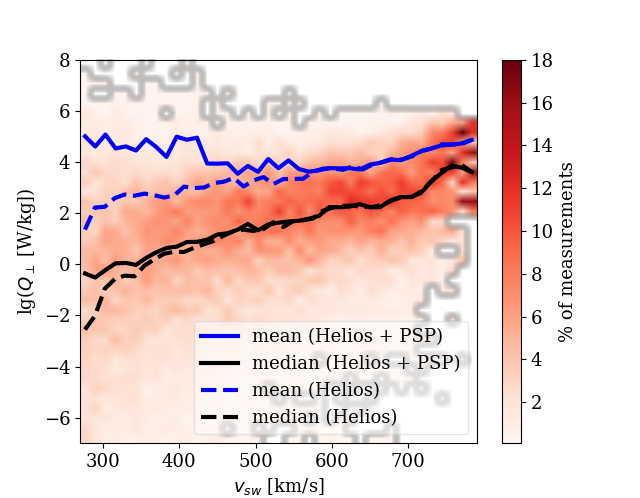}{0.49\textwidth}{(a)}
          \fig{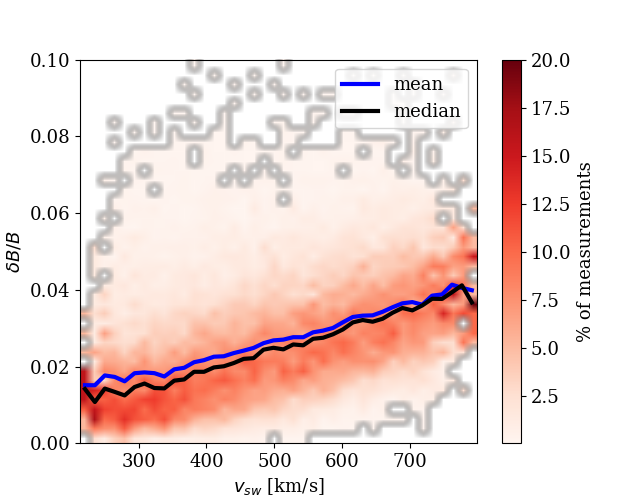}{0.49\textwidth}{(b)}
          }
    \caption{Values of the perpendicular SH rate $Q_\perp$ (a) and $\delta$ (b) are plotted against the solar wind bulk velocity (only low-$\beta$ plasma measured by both Helios and PSP is considered). The general logarithmic increase of $Q_\perp$ with $v_{sw}$ is notable, with values being significantly more spread for lower solar wind speeds. As influence of PSP measurements is visible only for $v_{sw}<600$ km/s, mean and median values from Helios results only are plotted separately as dashed lines to demonstrate the trend similar to the one shown on panel (b). Data in all of the panels shown is column normalized inside each velocity bin.}
    \label{fig:v_sw_dependencies}
\end{figure*}

In order to compare the measured $Q_\perp$ values with the total perpendicular heating of the solar wind protons, we estimate the total empirical heating rate $Q_{\perp emp}$ using Equation \ref{eq:Q_emp}. As explained in Section \ref{sec:Method}, all results are organized in bins divided by speed and radial distance and shown in different colors on Figure \ref{fig:Empirical_Heating_Rates}. Weighted averages that take into account the number of measurements in each of the velocity bins, as well as their uncertainties, are shown by black dots. These results are compared with calculated SH rates using Equations \ref{eq:Heating_Rate_low_beta} and \ref{eq:Heating_Rate}, marked by red and blue dots (PSP) and squares (Helios), respectively. The black dashed lines represent a simple model used by \citet{Bourouaine_2013}, based on evaluating Equation \ref{eq:Q_emp} at $r = 0.29$ au and extrapolating the result using radial trends of B, $v_{sw}$ and $v_{t\perp}$ measured by Helios, while the dashed red line shows the two-fluid model developed by \citet{Chandran_2011}, which is expected to be increasingly more accurate closer to the Alfv{\' e}n point. Predictions based on Helios measurements for fast and slow solar wind, filtered as $v_{sw} < 400$ km/s \citep{Hellinger_2013} and $v_{sw} > 600$ km/s \citep{Hellinger_2011} are shown by blue and green dash-dotted lines. For all of the four data sets shown, we note that the SH rates are comparable with the total estimated heating rates at radial distances covered by PSP results, while for $r>0.3$ au covered by Helios measurements, SH rates are lower than total heating rates, both measured and predicted by models, except for the results of \citet{Hellinger_2013}, which are relevant to slow wind only.

That the radial trend of total heating is significantly stronger in the fast solar wind is well known, being $Q \sim r^{-1.8}$ \citep{Hellinger_2011}, compared to $Q \sim r^{-1.2}$ for the slow wind \citep{Hellinger_2013}, leading to the correlation of the solar wind bulk velocity and temperature. This difference is even more visible for SH results, where MKB19 found $Q_\perp \sim r^{-3.1}$ for the fast wind in Helios results. Figure \ref{fig:v_sw_dependencies} (a) displays dependence of the low-$\beta$ $Q_\perp$ on the solar wind bulk velocity, where $Q_\perp$ increases for two orders of magnitude as $v_{sw}$ changes from 280 to 750 km/s. This trend is a direct consequence of a linear $\delta - v_{sw}$ correlation in Figure \ref{fig:v_sw_dependencies} (b), which implies that this significant increase in $Q_\perp$ originates from the intensity of turbulent fluctuations. 

\begin{figure}
\gridline{\fig{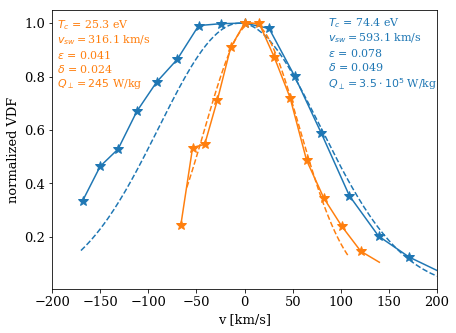}{0.49\textwidth}{(a)}
          \fig{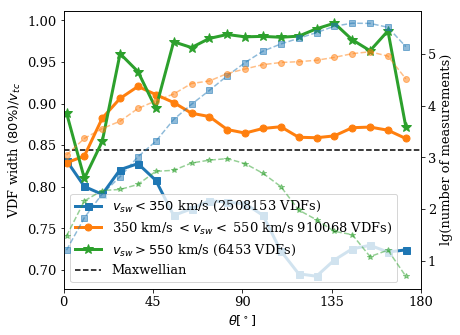}{0.49\textwidth}{(b)}
          }
    \caption{(a) Example of normalized measured reduced VDFs in the plasma frame, both perpendicular to the magnetic field, at radial distance of $r = 43.86 R_\odot$ (0.205 au) on November 2, 2018, 16:05 (orange) and November 9, 2018, 15:10 (blue). Maxwellian distributions with core temperatures provided by the instrument team are given by dashed lines. Panel (b) shows mean values of the VDF width at $80\%$ of the maximum value divided in 20 $\theta$ angle bins (of $9^\circ$ each), normalized to the core temperature $v_{tc}$. The mean VDF width is notably increasing with the solar wind speed for $\theta \sim 90^\circ$, which indicates the presence of flat-top distributions. Dashed lines map to the right-hand ordinate axis of panel (b) show number of VDFs sampled per angle and velocity bin, demonstrating that PSP has dominantly observed the slow solar wind, while statistics for the fast wind in radial and antiradial direction is poor.}
    \label{fig:normalized_VDF}
\end{figure}

%\gridline{\fig{mean_norm_VDF_slow.png}{0.49\textwidth}{(c)}
          %\fig{mean_norm_VDF_fast.png}{0.49\textwidth}{(d)}
          %}
%Mean values of the VDF in the plasma frame, normalized by the maximum value and fitted core thermal velocity $v_{tc}$, are shown for (c) $v_{sw} < 350$ km/s and (d) $v_{sw} > 550$ km/s solar wind. Flattening of the central part of the distribution is notable when angle between the bulk velocity and the magnetic field is larger than $30^\circ$ in the case of fast solar wind.

Finally, we perform a comparison of the SH values with the proton VDF shape.
Diffusion calculations of the SH dominated plasma by \citet{Klein_2016_ApJ} showed that thermal protons moving slowly in the perpendicular direction are more significantly affected by SH compared to protons with larger velocities in the perpendicular tail of the distribution. This behavior implies that protons with lowest perpendicular velocities are the ones gaining the largest amount of energy. The described effect is dominant only until particles reach velocities of the order of $v_t$, where their orbits become smoother (less stochastic) and SH from ion-scale fluctuations becomes less effective, resulting in a flat-top shaped reduced perpendicular distribution. In panel (a) of Figure \ref{fig:normalized_VDF}, we show example of a flat-top shaped reduced VDF measured in the fast solar wind (blue dotted solid line) perpendicular to the magnetic field. As the routine algorithm used by the instrument team assumes Maxwellian core, the quality of the fit (blue dashed line) is poor and the obtained core temperatures might be overestimated; see Figure 3 of \citet{Klein_2016_ApJ}. We see that the flat-top distribution corresponds to significantly stronger turbulent fluctuations compared to intervals with Maxwellian-like VDF measured in the slow wind (orange lines), with twice as large $\delta$ and $\epsilon$ parameters. The $Q_\perp$ calculated using Equation \ref{eq:Heating_Rate_low_beta} (both measurements are from low-$\beta$ streams and the same radial distance) is three orders of magnitude larger for the case of a flat-top core VDF than for a Maxwellian one. 

In the same work, \citet{Klein_2016_ApJ} fitted their flat-top VDFs with a modified Moyal distribution \citep{Moyal_1955}, which is characterized by decreased values of excess kurtosis $\kappa \approx -0.8$ compared to the Maxwellian value of $\kappa = 0$. However, this metric is only useful for analytic distributions spanning the entire velocity space. Even though a modified kurtosis can be derived for distributions that cover finite, but constant velocity range, our measured VDFs cover velocity ranges that change from distribution to distribution, depending on which of the detector velocity (voltage) channels are not polluted by the instrument noise.
These range changes affect the calculated kurtosis, confusing the physical interpretation of the values. Therefore, we track the systematic appearance of flat-tops by establishing a more direct criterion. We calculate the full width of the VDF at $80\%$ of its maximum height. This value is large enough to avoid influence from the proton beam or $\alpha$ particle populations. The width is normalized by the thermal velocity of the fitted core distribution $v_{tc}$, to both avoid the effects of the $v_{sw} \sim v_t$ correlation and to enable comparison with an equivalent Maxwellian distribution $f_{max}(v)$, which has width $w$ at $80\%$ of intensity
\begin{equation}
w[f_{max}(v)] (80\%) = 2 v_{tc} \sqrt{\mathrm{ln} \frac{1}{0.8}} \approx 0.845 v_{tc}
\label{eq:VDF_Max_Width_80}
\end{equation}
where $\mathrm{ln}$ is the natural logarithm. Results are shown on Figure \ref{fig:normalized_VDF} (b), where the solar wind is separated in three bins - slow, intermediate and fast. Slow wind results are below the referenced Maxwellian value, which is the effect of a finite measurement resolution combined with the generally colder, slower wind\footnote{As the SWEAP instrument is designed for maximum efficiency at the PSP closest perihelion ($r \sim 10 R_\odot$), measurements during encounters 1 and 2 were collected at the 'outer limits' of SPC's characteristics, while future observations are expected to have both signal to noise ratio and VDF resolution around the maximum significantly improved \citep{Kasper_2015,Case_2019_SPC}}. Importantly, the normalizes widths of VDFs systematically increase with solar wind speed, which implies that they also increase with measured values of $Q_\perp$ (Figure \ref{fig:v_sw_dependencies} (a)). This effect is visible when a reduced VDF is measured in the direction perpendicular to $\mathbf{B}$ for $v_{sw} > 550$ km/s, as expected from theoretical predictions \citep{Klein_2016_ApJ}. Increased widths that appear for low values of $\theta$ in the slow wind originate from measurements where the VDF maximum appears just next to the noisy low-energy channels. However, radial and anti-radial orientation bins are not covered with sufficiently strong statistics (green dashed line on panel (b)), so measurements from future PSP encounters will be required to examine these in more detail.

\section{Discussion}
\label{sec:Discussion}

The results shown in Section \ref{sec:Results} provide the first insight into the solar wind perpendicular heating rate measurements for radial distances in the range $0.16 - 0.25$ au. As such, we can test previously measured trends from larger distances and theoretical predictions.

It is particularly worth emphasizing that radial trends found in MKB19, which predict SH to be a significant fraction of the total solar wind heating closer to the Sun, are sustained at the radial distances covered by the first two PSP encounters. MKB19 estimate SH rate to be relatively insignificant at 1 au (with model parameters obtained also being in accordance with results of \citet{Vech_2017}), but having strong radial scaling as $Q_\perp \sim r^{-2.5}$, compared to the total heating rate scaling of $Q_{\perp emp} \sim r^{-2.1}$. This difference is even more significant in the fast solar wind, being $Q_\perp \sim r^{-3.1}$ against $Q_{\perp emp} \sim r^{-1.8}$, measured by \citet{Hellinger_2011}. Analysing results shown on Figure \ref{fig:Empirical_Heating_Rates}, we can compare the estimated total solar wind heating rates with the estimated SH rates. Total heating rate is shown by black dots, while the results obtained by the same method for larger radial distances are given by dashed line provided by \citet{Bourouaine_2013}. These values are either comparable or smaller than the estimated SH rates (red and blue circles) for $r \lesssim 0.2$ au, with even low-$\beta$ measurements becoming comparable to the empirical heating rates estimated using Equation \ref{eq:Q_emp}\footnote{The solar wind during the inbound phase of encounter 1 has been determined to be co-rotating with a small coronal hole \citep{Bale_2019_Nature,Badman:prep}, and it has been reported \citep{Bale_2019_Nature} that the turbulent fluctuations are very low in this period, which causes the measured SH values to be significantly lower then those obtained from MHD models, as shown on Figure \ref{fig:Empirical_Heating_Rates} (upper left).}. Of course, SH being higher than total heating rate is unphysical result, but is still within range of relatively large measurement uncertainties, which are approximately one order of magnitude for both sets of values. On the other hand, SH rates estimated from Helios measurements are either systematically lower or comparable with the observational results given by dashed black line. Considering these results and findings from MKB19 and \citet{Vech_2017}, we conclude that SH is playing in increasingly important role in the solar wind heating as the radial distance is decreasing, and possibly becoming a dominant heating mechanism in the near-Sun solar wind. 

For further testing of this statement, we were able to obtain measurements of flat-top shaped distributions, as shown in Figure \ref{fig:normalized_VDF}. This result is in accordance with predictions of the model that assumes dominant SH \citep{Klein_2016_ApJ}. Figure \ref{fig:normalized_VDF} (a) gives an example of a flat-top distribution at $r = 0.205$ au during the encounter 1 outbound phase, with the SH rate that contains a significant fraction of total heating predicted both by models and the total heating calculation (Figure \ref{fig:Empirical_Heating_Rates} (upper right)). Even though the statistics of fast wind measurements are poor, we found that this VDF shape in the direction perpendicular to $\mathbf{B}$ is commonly measured in the fast wind, as shown in Figure \ref{fig:normalized_VDF}, panel (b). However, although our results suggest that the SH rate is similar to the estimate of the total heating in the fast wind and visual inspection of the measured VDF shows the flat-top shape, averaged values of normalized VDFs in the fast wind shown on Figure \ref{fig:normalized_VDF} (b) imply that a modified Moyal distribution is not fully developed. The observed averaged distributions have width at $80\%$ of maximum of approximately $w \approx v_t$, while the Moyal, as predicted by \citet{Klein_2016_ApJ}, has the notably larger value of the same parameter at $w \approx 1.3 v_t$. Due to limitations of the data set,  at this point is not possible to distinguish if this discrepancy between theory and observations is caused by mechanisms other than SH notably participating in the proton diffusion in velocity space (and, consequently, heating of the solar wind) or by SH being the dominant heating mechanism, but the modified Moyal VDF shape is not sustainable in realistic solar wind conditions. The recent argument in \citet{Isenberg_2019_ApJ} suggests that extreme flat-top distributions can be suppressed by a quasi-linear ion-cyclotron anisotropy instability for large temperature anisotropy in very low-$\beta$ plasma. In the measured streams, we have $T_\perp / T_{||} \leq 5$ on more than $90\%$ of the intervals \citep{Huang:prep}, so this effect is not expected to dominate at the PSP distances, but rather be a competing process with the perpendicular particle diffusion described by \citet{Klein_2016_ApJ}, with importance depending on multiple parameters. A possible interpretation of the observed shape is that the ion-cyclotron instability could be important close to the Sun and we observe more Maxwellian-like VDF shape after strong instabilities have been saturated.

An important property of the results shown on Figure \ref{fig:Q} is that $Q_\perp$ varies over 10 orders of magnitude with only $\sim 10-15\%$ of values being above average, and that there is a two order of magnitude difference between mean and median. This is seen in both Helios and PSP measurements. These features have been discussed in detail in MKB19, where it was argued that $\mathrm{lg}(Q_\perp)$ and $\delta B$ follow a Gaussian distribution with an addition of a broad low-energy tail. This tail is associated with the pristine, low-turbulent amplitude solar wind and negligibly contributes to mean heating values. Also, due to strong $v_{sw} - \delta$ correlation visible on Figure \ref{fig:v_sw_dependencies} (b), the spread in $Q_\perp$ is much less if slow and fast wind are treated independently. Finally, effects of highly intermittent turbulence are not taken into account in this work. Strong intermittency can significantly increase the SH rate, according to recent theoretical results \citep{Mallet_2019}. Case studies of intermittent intervals is a separate topic for future work.

Along with discussing these results, we need to be aware of all of the limitations of both methods and measurements used. The PSP high resolution data from encounters 1 and 2 covers only two 11-day long encounter periods separated by a five-month span between the two closest approaches. Therefore, we don't expect a large, statistically complete set of results, but only a 'randomly picked' sample of solar wind plasma at $0.16 - 0.25$ au, almost $90\%$ of which is from the slow solar wind. In contrast, SH (and solar wind heating in general) is expected to be more effective in the fast solar wind, where our measurements are severely lacking. Measurements from few several hours-long streams at $r \sim 0.2$ au during E1 outbound phase are the rare exception. These fast wind results correspond to the maxima at Figure \ref{fig:Empirical_Heating_Rates} (upper right) for both stochastic and total heating, as well as to the illustrative example on Figure \ref{fig:normalized_VDF} (a). Wider insight into SH behavior in the slow and fast streams separately, analogous to the one performed for Helios data in MKB19, will be possible once the data set is enlarged during the following encounters. In investigating intervals near and in Alfvenic 'spikes' \citep{Horbury:prep}, increases in both fluctuation amplitudes and Poynting flux were observed near the spike boundaries \citep{Kasper_2019_Nature,Horbury:prep}. This relatively short (up to tens of seconds) timescale boundary effect is not expected to be important for the results presented here, as we use 10-minute averaged measurements, but properties of SH in different scale structures are a separate topic for future work. This kind of case study is, with current instrumentation, possible only for PSP and at 1 au, due to lower quality and resolution of Helios measurements at intermediate radial distances.

Combined with the uncertainties in measurements, it is important to note that the SH calculation method described in Section \ref{sec:Method} can only provide an order of magnitude estimates of $Q_\perp$ due to selected values of the model constants $c_1$, $c_2$, $\sigma_1$ and $\sigma_2$. In this work, we use values provided by test particle simulations for randomly phased Alfv\'en waves \citep{Chandran_2010,Hoppock_2019}. Values of $c_1$ and $c_2$ were also examined by \citet{Xia_2013} in test-particle simulations, but the with test particles interacting with RMHD turbulence rather than randomly phased waves, significantly lower values of $c_2 \sim 0.2$ were found. This value is used as an alternative in our model in Figures \ref{fig:Q} and \ref{fig:Empirical_Heating_Rates}, demonstrating that estimated low-$\beta$ SH rates are increased by more than an order of magnitude. No other estimates of $\sigma_1$ and $\sigma_2$ are available beyond those provided by \citet{Hoppock_2019}. Detailed discussion on constants is provided in MKB19. Finally, it is important to note that the $\sigma$ factor, which connects magnetic and velocity fluctuations, assumes Alfv{\' e}nic turbulence. If the turbulence is not purely Alfv{\' e}nic, this value could vary, which would produce similar effect to changing $c_2$ or $\sigma_2$ - a notable variation in $Q_\perp$ for high-$\beta$ plasma. Finally, we conclude that all the factors described here enable us to only determine an order of magnitude of the SH rate, in a similar fashion to the Helios data analysis in MKB19. On the other hand, simulation results \citep{Chandran_2010,Xia_2013} show that the values of model constants, even though unknown, are not very sensitive to plasma parameters, and are therefore still usable in determination of the radial trends related to SH, which was discussed above.

In the encounters to come, we will aim to further test these results with a larger sample of fast wind measurements at different radial distances. Once a more robust data set becomes available, a case study of these particular intervals, relying on the assumption that they are determined by SH as the only solar wind proton perpendicular heating mechanism, will provide useful constraints on the model constants and significantly reduce uncertainties of the SH rate amplitudes.

We expect an increased level of turbulence below the Alfv{\' e}n point, predicted by theoretical models \citep{Matthaeus_1999} and MHD simulations \citep{Perez_2013_ApJ,Chhiber_2018_ApJ,Chandran_2019_JPlPh}, to effectively drive SH. Resent results \citep{Kasper_2019_Alfven} indicate that the region of preferential minor ion heating \citep{Kasper_2017} terminates at the Alfv\'en surface. Whether SH is the preferential heating mechanism operating in this region will be determined by future PSP measurements within this region. With the data from the first two encounters, we argue that SH tends towards being the dominant solar wind proton heating mechanism as PSP approaches the Alfv{\' e}n point, especially in the fast solar wind streams. This is consistent with both Helios analysis and both approaches presented in this work \,-- determination of SH rate using theoretical formalism \citep{Chandran_2010,Hoppock_2019} and analysis of VDF shapes according to the predictions made by \citet{Klein_2016_ApJ}, as well as comparison of these results with the total heating rate estimate from PSP data and theoretical models \citep{Chandran_2011,Bourouaine_2013}. 

\acknowledgments

The SWEAP Investigation and this publication are supported by the PSP mission under NASA contract NNN06AA01C.
The FIELDS experiment was developed and is operated under NASA contract NNN06AA01C. 
M. M. Martinovi\'c and K. G. Klein were financially supported by NASA grant 80NSSC19K1390. B. D. G. Chandran is partially supported from NASA grants NNX17AI18G and 80NSSC19K0829. C. H. K. C. is supported by STFC Ernest Rutherford Fellowship ST/N003748/2.

%\bibliographystyle{aasjournal}
%\bibliography{Latex_Refs}

\end{document}